\documentclass[aps,twocolumn,groupaddress,showlabels,showrefs,amsmath,amssymb,pre,superscriptaddress, floatfix, colors]{revtex4}

\usepackage{graphicx}% Include figure files
\usepackage{dcolumn}% Align table columns on decimal point
\usepackage{bm}% bold math
\usepackage{amssymb}
\usepackage{multirow}
\usepackage{color}
\usepackage{hyperref}

\usepackage{graphicx}
\usepackage{amssymb}
\usepackage{amsmath}
\usepackage{amsfonts}

\begin{document}

\title{Active Brownian  equation of state: metastability and phase coexistence}

\author{Demian Levis}

\affiliation{Departament de F\'isica de la Mat\`eria Condensada, Universitat de Barcelona, Mart\'i i Franqu\`es, E08028 Barcelona, Spain}
\author{Joan Codina}
\affiliation{Departament de F\'isica de la Mat\`eria Condensada, Universitat de Barcelona, Mart\'i i Franqu\`es, E08028 Barcelona,   Spain}
\author{Ignacio Pagonabarraga}
\affiliation{Departament de F\'isica de la Mat\`eria Condensada, Universitat de Barcelona, Mart\'i i Franqu\`es, E08028 Barcelona, Spain}

\begin{abstract}
As a result of the competition between self-propulsion and excluded volume interactions, purely repulsive self-propelled spherical particles undergo a  motility-induced phase separation (MIPS).  
We carry out a systematic computational study, considering several interaction potentials, systems confined by hard walls or with periodic boundary conditions, and different initial protocols. 
This approach allows us to identify that, despite its non-equilibrium nature, the equations of state of Active Brownian Particles (ABP) across MIPS verify the characteristic properties of first-order liquid-gas phase transitions, meaning, equality of pressure of the coexisting phases once a nucleation barrier has been overcome and, in the opposite case, hysteresis around the transition as long as the system  remains in the metastable region. 
Our results show that the equations of state of ABPs quantitatively account for their phase behaviour, providing a firm basis to describe MIPS as an equilibrium-like phase transition.  
\end{abstract}

%\pacs{pacs}

\maketitle
Active matter made of self-propelled particles can be found in a wide variety of   contexts, both in  living and synthetic systems,  ranging from  cell populations to bacteria suspensions, animal groups  or colloidal artificial swimmers \cite{MarchettiRev}. 
The fundamental difference between active  and  'passive' matter made of thermally agitated constituents, is that the microscopic dissipative dynamics of the former breaks detailed balance and, as such, evolves out-of-equilibrium. The intrinsic out-of-equilibrium nature of active matter manifests strikingly in the presence of different kinds of  interactions. Self-propelled spherical particles accumulate in regions of space where their velocity decreases as a consequence of collisions (between particles or with external obstacles).
Simple models  that capture in a minimal way the competition between self-propulsion and steric effects, like the so-called Active Brownian Particles (ABP) model, have provided much insight into the generic behavior of such systems.  
For instance, at high enough densities and activities, a purely Motility-Induced Phase Separation (MIPS) generically takes place, leading to the coexistence of an active low density gas with a high density drop in the absence of attractive forces~\cite{TailleurCates2008, TailleurCatesRev, Fily2012, Redner2013,  Stenhammar2014, Levis2014}. 
This out-of-equilibrium transition is  reminiscent of  equilibrium liquid-gas de-mixing. It is thus tempting to   extend the thermodynamic description of  first order phase transitions in terms of, for instance, equations of state, to ABPs. However, this poses several fundamental difficulties since no thermodynamic variable is, in principle, well defined in this context. Much effort has been recently devoted to this question from a statistical mechanics perspective: the notions of effective temperature \cite{Loi2008, Palacci2010, Szamel2014, Levis2015, Marconi2015}  and chemical potential \cite{Dijkstra2016} have been introduced, and special attention has been paid to the notion of  pressure in active fluids \cite{Takatori2014,Yang2014, Solon2015, Ginot2015, Winkler2015,  Patch2016, Speck2016}. In particular, it has been  shown that an equation of state exists for isotropic ABP. %, with equality of pressure at coexistence 
% \cite{Solon2015} (though this property is not as robust as in equilibrium \cite{Solon2015bis}).

In spite of the above-mentioned efforts, the equality of pressure at coexistence implied by the very existence of an equation of state has not been observed so far (neither in experiments nor simulations) and the possibility to construct from the equations of state an effective thermodynamic description of MIPS is still a matter of debate.  
In order to move forward in our fundamental understanding of active systems, it is crucial to provide a consistent interpretation of pressure measurements that allows to confront them with the theory. % and therefore characterize their phase behaviour.
%Despite these recent works, the equations of state of ABP are still not fully understood and the possibility to construct from them an effective thermodynamic description of MIPS is still a matter of debate. 
%Within this spirit, i
% and  comparison with theoretical predictions. 
%In order to develop a well-grounded understanding of active systems, it is fundamental to provide a systematic way to interpret the equations obtained in experiments or simulations of model systems and be able to characterize the phase behaviour of active liquids from pressure measurements (and to compare the collected data with the theory). 
In this Letter we clarify this issue and provide a full description of the phase behavior of ABPs in terms of its equations of state. %of the pressure measurements across MIPS 
%that reconciles previous conflicting results.

%Neither the current experimental nor simulation data allows to fully confirm the theoretical scenario of pressure coexistence across MIPS: 
 %\cite{Takatori2014,Yang2014, Solon2015, Winkler2015,  Patch2016}. 
The equations of state of self-propelled Janus colloids have been experimentally measured in the absence of phase coexistence \cite{Ginot2015}. Several numerical studies have measured the pressure in systems undergoing MIPS, but the way to interpret the results raises some serious conceptual problems \cite{Takatori2014, Winkler2015,  Patch2016}.
%The pressure in systems undergoing MIPS has been measured in several numerical studies. 
For instance, an abrupt pressure drop at the vicinity of MIPS
%, occurring at densities well above the coexistence ones,  
has been reported recently and it has been argued that it constitutes a distinctive feature of this transition, hindering the analogy with equilibrium phase separation \cite{Winkler2015,  Patch2016} and in apparent contradiction with  theoretical predictions \cite{Solon2015, Solon2016}. 
A pressure loop generically appears in % $NVT$ simulations of 
 phase separating finite systems in equilibrium,   
%(due to the interface tension in finite size systems)
 an effect that can be suppressed by the Maxwell construction to extract the thermodynamic behavior. % of the system in the thermodynamic limit. % in order to obtain the coexistence densities and pressure. 
However, this construction is violated for ABP \cite{Solon2015,Solon2016}, so there is no direct way to understand the phase diagram of the system from the simulated equations of state.
% in order to construct the binodals. 
%Moreover, there are some inconsistencies in the simulation results reported so far: either the first virial coefficient varies as we move in the  coexistence region \cite{Takatori2014}, or, on the contrary, remains constant \cite{Solon2015, Winkler2015, Patch2016}. 
%Therefore, 
%a  fundamental understanding of the simulation results is still lacking since 
%the %equations of state
 %measurements obtained so far do not allow to characterize the phase behaviour of ABP and, in particular, the nature of MIPS. 
 %The  existence of an equation of state implies the equality of pressure of coexisting phases, a fundamental result that has not been properly confirmed  yet. 

We show here that the equations of state of ABP across MIPS are consistent with the first-order phase transition scenario and allow to characterize its phase behaviour, providing a consistent thermodynamic interpretation of the numerical data. % in agreement with the theory. 
%Our aim here is to show to what extent one can use the equations of state obtained from simulations to describe the phase behaviour of ABP.
% and discuss   the analogies between MIPS and first order liquid-gas transitions. 
%We  %perform numerical simulations of a model system, and 
%show that the equations of state agree with the first-order phase transition scenario. 
%By changing the preparation protocol and the topology of the system, 
We bring out the existence of a metastability region%between the onset of phase separation and the binodal
: an \emph{hysteresis} around the coexistence pressure is found %in the equations of state 
as the system is quenched to the coexistence region from `below' (from a homogeneous state) or `above' (from a phase separated state) MIPS. We analyze both open and confined systems to show that all the conflicting results 
%pressure drop (and the invariance of the first virial coefficient) 
found in previous simulation studies %\cite{Solon2015, Winkler2015, Patch2016} 
are due to the presence of a large nucleation barrier that can be easily bypassed by %initializing the system in the coexistence region or 
including a nucleation core (a wall). Then one can generate %numerically
a motility induced phase separated state %arbitrarily close to the binodal, i.e. 
at densities well below those previously reported. 
%, in agreement with a thermodynamic picture. 
To show these results, we perform constant-density simulations ($NVT$ ensemble) 
 of  Active Brownian Hard-Disks and  ABP interacting with a WCA potential. 
%We thus find that there is no minimal packing fraction fro MIPS.   
%defining thermodynamic variables in active systems. 

%Similarities between active and passive systems. Phase separation. Use of thermodynamic concepts. Simulations of equations of state Brady, Gompper, Marchetti, give contradictory results: pressure jump and virial coefficients which do not vary inside the coexistence region. This is odd. Phase separation only observed above a threshold value that has been estimated using kinetic models (Redner Baskaran). 

%Here we show that there is a region of metastability, with hysteresis like in usual first-order transitions. The pressure jump is due to the fact that there is a huge nucleation barrier to overcome when the system is quenched from the homogeneous phase to the inhomogenous one. We also discuss the role of the particle stiffness in the phase coexistence.    

%Breve resumen de los resultados mas relevantes.

To be specific, we consider $N$ disks of diameter $\sigma$ in a 2$d$ volume $V=L_x\times L_y$, with  $\phi=\pi\sigma^2N/4V$ the packing fraction. The model is defined by the following  equations of motion for each particle at position $\boldsymbol{r}_i=(x_i, y_i)$ and with orientation $\boldsymbol{n}_i(t)=(\cos\theta_i, \sin\theta_i)$:
\begin{equation}\label{EQ:langevin}
\dot{\boldsymbol{r}}_i= v_0\boldsymbol{n}_i +\mu \boldsymbol{F}_{i} +\sqrt{2D_0} \boldsymbol{\xi}_i\, ,  \dot{\theta}_i=\sqrt{2D_{\theta}} \nu_i
\end{equation}
where $\boldsymbol{\xi}_i$  and $\nu_i$ are zero-mean unit-variance Gaussian noises, and $v_0$ is a constant self-propulsion velocity. 
The  force $\boldsymbol{F}_{i}$ acting on particle $i$ comes from inter-particle interactions, $\boldsymbol{F}^{int}_{i}=-\frac{1}{2}\nabla_i \sum_{j\neq i}u(r_{ij})=\frac{1}{2}\sum_{j\neq i}\boldsymbol{f}_{ij}$,  and external potentials, $\boldsymbol{F}^{ext}_{i}=-\nabla_i w(r_{i})$. 
The units of  length and time are given by $\sigma$ and $\tau=D_{\theta}^{-1}$, respectively, and fix $D_{\theta}=3D_0/\sigma^2$. 
Together with the packing fraction, we identify two dimensionless parameters that control the phase behaviour of our system: the P\'eclet number, $\mbox{Pe}={v_0}/{\sigma D_{\theta}}$, which quantifies the strength of self-propulsion, and the effective particle stiffness $\Gamma=\frac{\epsilon \mu}{\sigma v_0}$, %mbox{Pe}$, 
which naturally arises from the potential energy scale $\epsilon$ (in units of $k_B T=\mu/D_0=1$). This % non-dimensional 
parameter quantifies to what extent particles become effectively softer as their activity increases, an effect  surprisingly disregarded in the literature. 

 In order to   focus on the essential features determining the phase behavior of the system, we compare  a suspension of Active Brownian Disks with a Weeks-Chandler-Andersen potential (AB-WCAD): 
$u(r)=4\epsilon\left[ \left(\frac{\sigma}{r}\right)^{12}-\left(\frac{\sigma}{r}\right)^6+\frac{1}{4}\right]$ with a upper cut-off at $r=2^{1/6}\sigma$
% ***in the range $\Gamma=3, ..., 100$ 
and such that $\Gamma>0.03$; with one composed by infinitely hard  active particles, i.e.  $u(r)=0$ if $r>\sigma$ and $u(r)=\infty$ otherwise  \cite{SM}. The hard sphere fluid is the most studied model in liquid theory: 
 it was the first fluid to be simulated with molecular dynamics \cite{Alder1957} and several analytical developments can be carried to predict its thermodynamic properties \cite{HansenBook, Baxter1968,Carnahan1969}. 
In both cases  we carry out BD simulations to study their evolution  in the absence of any external forcing. It is not straightforward to deal with the singular nature of hard core interactions in Brownian dynamics (BD) simulations, %which are 
commonly used in the context of ABP. 
 %\cite{Fily2012, Redner2013, Yang2014,  Stenhammar2014, Solon2015, Winkler2015,  Patch2016}. 
We therefore use a variant of  Even-driven Brownian Dynamics (ED-BD)  for hard spheres \cite{Scala2007} that accounts for self-propulsion \cite{Ni2013}\footnote{The constraint that two particles cannot overlap can be implemented by simply discarding displacements that do not obey this condition  - this corresponds to the standard Monte Carlo (MC) procedure used in \cite{Levis2014} - or by performing elastic collisions when two particles are at contact, as we do here. The MC dynamics with persistence used in \cite{Levis2014} gives rise to clusters with self-limiting size, while using  ED-BD  we find clusters that coarsen by MIPS, as generically observed in previous BD simulations of  ABP. }.
 %\cite{Fily2012, Redner2013, Yang2014,  Stenhammar2014, Solon2015, Winkler2015, Patch2016}. 
 %Indeed, as we shall discuss in more detail below,  
%Active Brownian Hard Disks (AB-HD) with ED-BD can be thought of as the limit of infinite stiffness of the standard ABP model with a continuous interaction potential.  

\begin{figure}
\centering
\includegraphics[scale=0.65,angle=0]{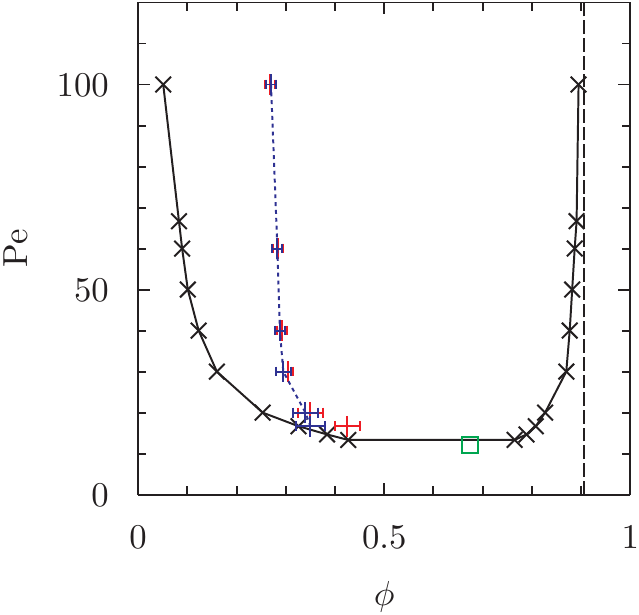}\, \includegraphics[scale=0.65,angle=0]{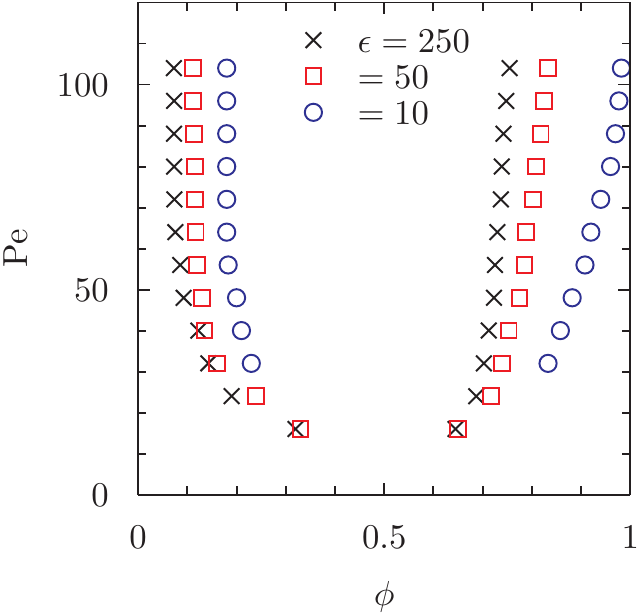}
\caption{ Pe-$\phi$ phase diagram of ABP. Left: AB-HD. Black symbols show the coexisting densities defining the binodals. Red points indicate the onset of MIPS, measured from the %peak of 
the second moment of the fraction of particles in the largest cluster (see \cite{SM} for details). Blue points correspond to the location of the pressure drop in open systems (see below). We identify a critical point at  Pe$_c\approx12$ and $\phi_c\approx 0.674$ (shown by a green symbol). The vertical dotted line indicates the densest possible packing of hard disks $\phi_{\text{cp}}=\pi/(2\sqrt{3})$.
Right: AB-WCAD for different stiffness. The symbols show the two binodals $\phi_{\text{low}}$ and $\phi_{\text{high}}$.
}
\label{fig:phd}
\end{figure}

The phase diagrams shown in Fig. \ref{fig:phd} can be readily constructed from the coexisting densities. 
%In the following we show  that the equations of state are consistent with this phase diagram, making the connection between the mechanical properties of ABP and its phase behaviour, just as in equilibrium thermodynamics. 
%, as it has been done previously \cite{Redner2013}
%We construct the phase diagrams Fig. \ref{fig:phd} from the distribution of density at coexistence. 
ABP with different stiffness exhibit the same qualitative phase behaviour: above a critical P\'eclet number they undergo MIPS if its average density is large enough. Increasing the potential stiffness one observes both a decrease of the coexisting packing fractions and an increase in the structure, as depicted in Fig. \ref{fig:nuc} (d). This feature implies that the structure of the high-density phase is more compact for stiffer potentials. In the limit of hard disks, the high-density branch of the binodal quickly saturates at close packing, indicating that the competition between self-propulsion and excluded volume generates a nearly perfect hexagonal crystal at high enough activities, as shown in Fig. \ref{fig:nuc} (d) (e.g. for Pe$=60$, the average density of the condensate is $\phi_{\text{high}}=0.98\, \phi_{\text{cp}}$, in coexistence with a low density phase at $\phi_{\text{low}}= 0.089$). 

We identify the pressure, $P$, by an extension of the virial theorem that accounts for  active forces \cite{Winkler2015, Falasco2016}. After projecting eq.~(\ref{EQ:langevin}) on $\boldsymbol{r}_{i}$ and averaging over 
%independent realizations of 
the noise, we get
%\begin{equation}\label{EQ:ProjectedWall}
$P=  \rho k_B T  +   P_{int}+ P_{a} $, 
  %\end{equation}
 where  $\rho k_B T$ is the thermal ideal gas pressure, 
\begin{equation}\label{eq:pi}
P_{int}= \frac{1}{4V} \sum_i  \sum_j \langle \boldsymbol{f}_{ij} \cdot (\boldsymbol{r}_i-\boldsymbol{r}_j)\rangle
\end{equation}
is the standard virial expression of the collisional pressure, %\footnote{For infinitely hard spheres, use Adler procedure, count collisions. }, 
and 
\begin{equation}\label{eq:pa}
P_a=  \frac{v_0}{2 \mu V}\sum_i \langle \boldsymbol{n}_i \cdot \boldsymbol{r}_i\rangle =  \frac{\rho v_0 v_{\phi}}{2 \mu D_{\theta}}\, , \ v_{\phi}= \frac{1}{N}\sum_i  \langle\boldsymbol{n}_i \cdot \dot{\boldsymbol{r}}_i\rangle
\end{equation}
is the active contribution to the pressure or 'swim' pressure, which can be mechanically interpreted as a body force \cite{Takatori2014,Takatori2015, Yang2014}. This yields an ideal gas law $P_0= \rho k_BT[1+{ v_0^2}/{(2 D_0 D_{\theta}})]$. The total pressure $P$ defined this way is a state function for spherical ABP \cite{Solon2015} and its definition can be extended to unconfined systems with PBC where $P$ is equal to the internal bulk pressure \cite{Winkler2015}. 

Following usual practice, we study the phase behaviour of ABP by letting the system evolve from a homogenous initial state at some density $\phi$  towards its stationary state corresponding to a given value of $\text{Pe}$. This procedure corresponds to a quench from $\text{Pe}=0$, below MIPS, to $\text{Pe}>0$.  However, in order to gain insight in the intrinsic nature of the pressure, we compare the values obtained for a periodic system and one confined in  the $x$-direction by two walls at $x=0$ and $x=L_x$, and  periodic boundary conditions (PBC) in the $y$-direction \footnote{We use, for convenience, an external confining potential $w(x)=\left[\left(\frac{1}{x}\right)^{12}+\left(\frac{1}{L_x-x}\right)^{12}\right]$. Then the wall-particle force acts along the $x$-axis:  $\boldsymbol{F}^{ext}(\boldsymbol{r}_i)=\boldsymbol{F}^{ext}(x_i)=-\partial_{x_i} w(x_i) \boldsymbol{u}_x$, where $\boldsymbol{u}_x=(1,0)$ is the base vector.}. 

\begin{figure}
\centering
\includegraphics[scale=0.6,angle=0]{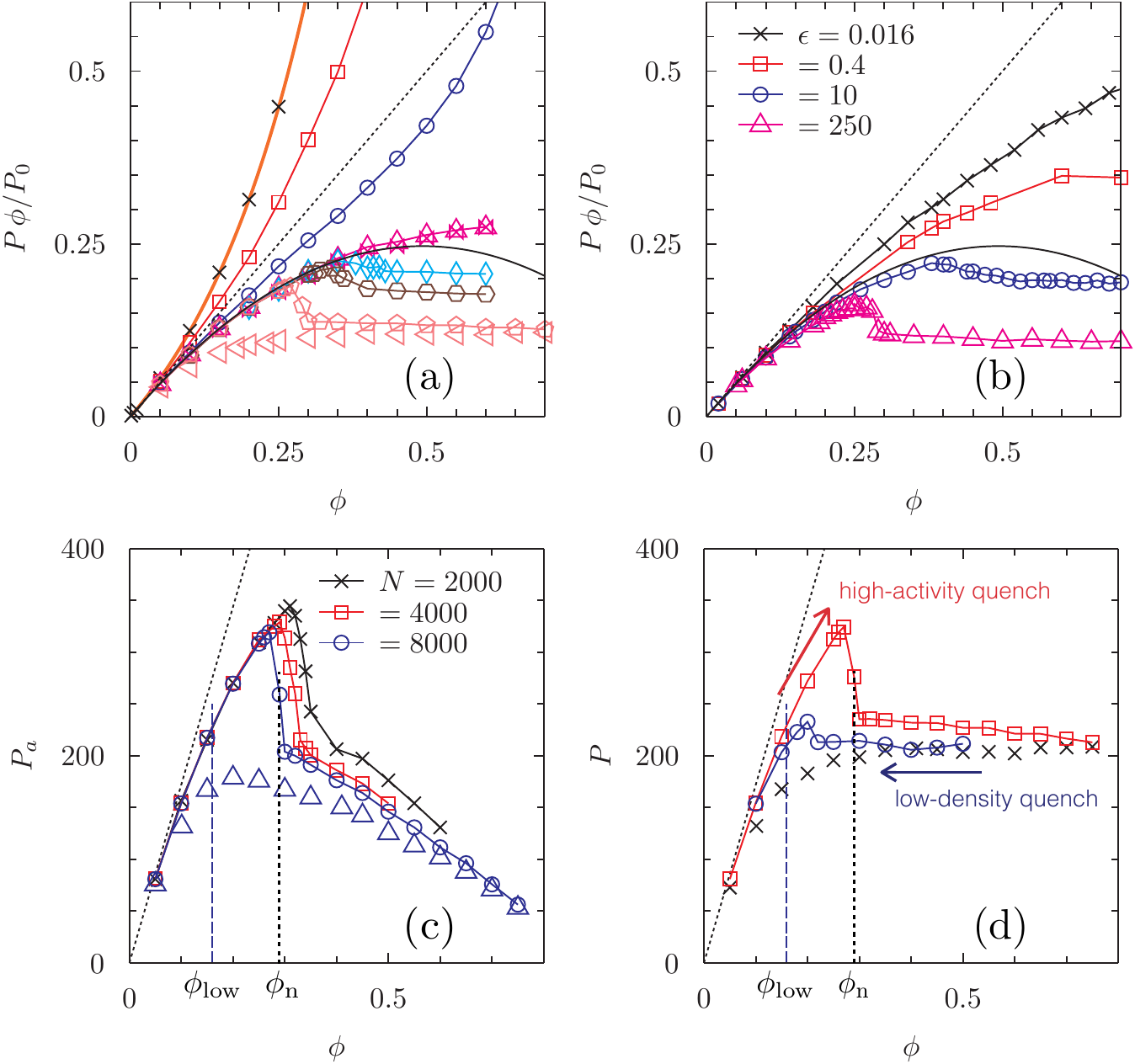}
\caption{(a): Total pressure times $\phi$ normalized by the ideal gas pressure $P_0$ of AB-HDs for Pe=0 (black), 1 (red), 3 (blue), 10 (purple), 16.7 (cyan), 20 (brown) and 30 (pink). 
%$N=4000$. 
The continuous (orange) line corresponds to the Carnahan-Starling equation of state; the dotted one is the ideal gas limit.   The continuous black line indicates the first virial correction to the ideal gas
  %, i.e. $P\phi/P_0=\phi(1+B_1\phi)$, 
with $B_1=-1$. (b): Total pressure of AB-WCADs of different stiffness %$N=1250$ 
at fixed Pe$=100$. The continuous and dotted lines are the same as in (a). (c): Active pressure for systems of AB-HDs of different size for Pe$=30$. (d): Hysteresis in the equations of state of AB-HDs for Pe$=30$: %Here $N=8000$ and Pe=$30$.
The red points are obtained using the (usual) protocol of quenching the system from an homogenous state ($\text{Pe}=0$) to $\text{Pe}>0$ ; the blue points were obtained by quenching the system from a phase separated state ($\phi=0.50$, $\text{Pe}=30$) to lower densities.  Black points show the equation of state in the confined system. The vertical dotted line in (c) and (d) indicates the coexistence density $\phi_{\text{low}}$. 
}
\label{fig:eos}
\end{figure}

\begin{figure}
\includegraphics[scale=0.6,angle=0]{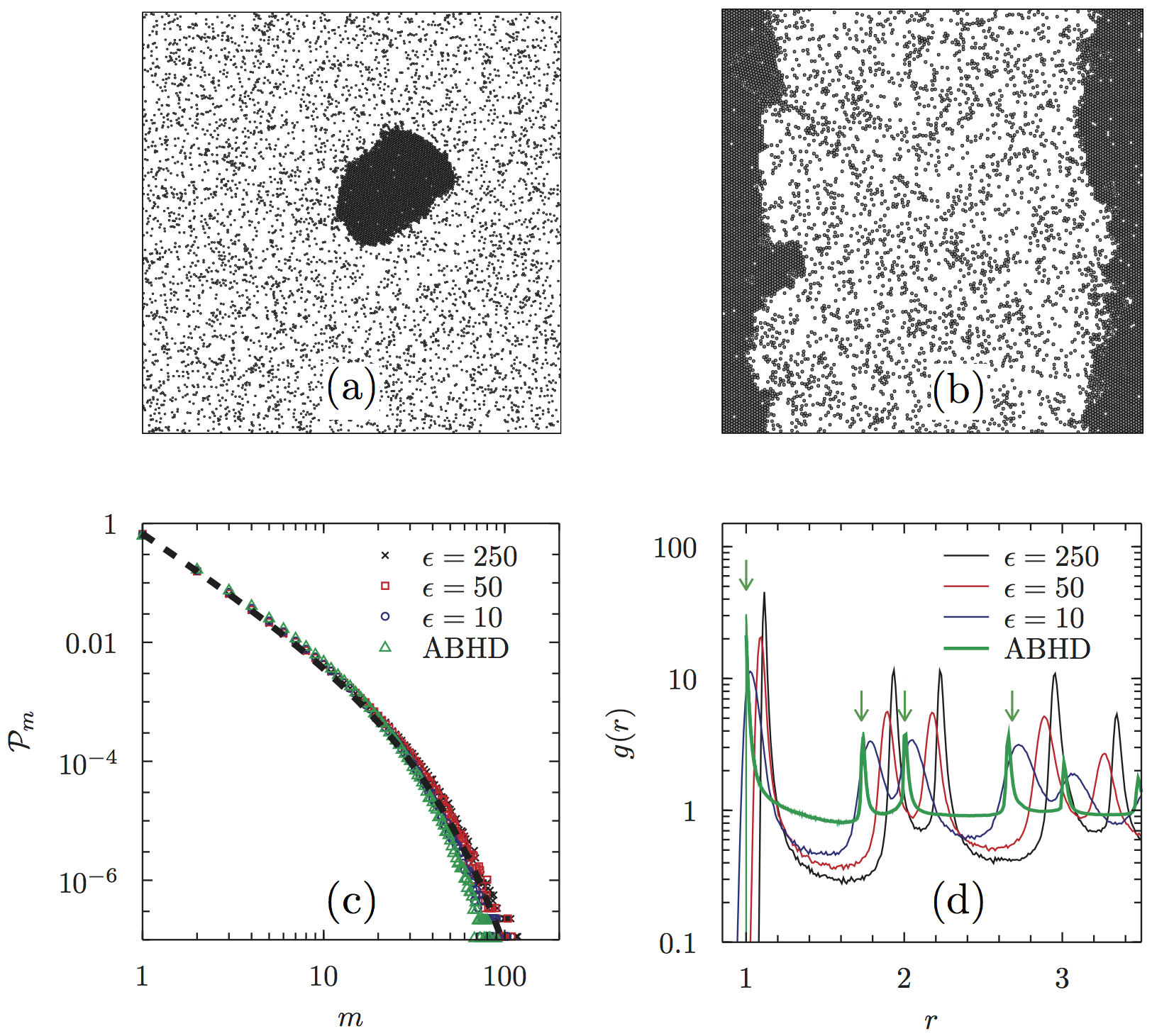} 
\caption{(a) Snapshot of an open system of AB-HD with PBC at $\phi=0.12$ and Pe=$60$ obtained from a low-$\phi$ quench. (b) Snapshot of AB-HD in the confined geometry for $\phi=0.30$ and Pe$=30$.  (c) Cluster size distribution for ABP, both AB-HD ($\epsilon\to\infty$) and AB-WCAD,  with PBC at $\phi=20$,$\text{Pe}=30$ for several stiffness. For comparison, we shown in dotted lines: $\propto \exp(-m/m^*)/m^2$ with %$A=0.71$ and 
$m^*=14.5$.  (d) Pair correlation function at fixed Pe$=30$ and $\phi=0.35$ for several potentials. The arrows indicate the peaks  at $r=1$, $\sqrt{3}$, $2$ and $1+\sqrt{3}$ corresponding to an hexagonal packing.}
\label{fig:nuc}
\end{figure}

The $P(\phi;\,$Pe$)$ equations of state computed numerically using eq. (\ref{eq:pi}-\ref{eq:pa}) in confined and  PBC geometries, following  \emph{high-Pe quenches} at fixed $\phi$, are shown in Fig. \ref{fig:eos} (a-c).  If the system does not phase separate, 
for  Pe$<$Pe$_c$, the total pressure of the AB-HD fluid with PBC coincides with the pressure computed in confined conditions (see Fig. \ref{fig:eos} (a)).
 %{\bf IGN: una precision: en presencia de la pared incluimos el potencial confinante  en el calculo del virial y obtenemos la presion del sistema promeidando sobre tod ala caja, correcto? tal como est aescrito, parece que calculemos la presion que las ABPs ejercen sobre la pared de forma mecanica}. 
At higher activities, for PBCs  the pressure drops abruptly  in the vicinity of MIPS \cite{Winkler2015, Patch2016}, while in the confined system the pressure displays a  monotonic increase with  density and remains  roughly constant in the coexistence region. 
%shows a behaviour closer to what it is expected for an infinite system in phase coexistence.   
As shown in Fig. \ref{fig:eos} (a-b), the pressure jump becomes more pronounced as the activity and the stiffness of the particles increases (note the similarity between %the equations of state for 
AB-HD with $\text{Pe}=30$ and AB-WCAD with $\epsilon=250$ and $\text{Pe}=100$). 
%\textcolor{blue}{[Say already here why? Nucleation barrier, larger gap between the binodal and the spinodal and more crystalline cluster, i.e. larger surface tension ? ]}
The formation of a macroscopic drop, due to MIPS, coincides with the location of the pressure drop, $\phi_{\text{n}}$ (see \cite{SM}). Particles are dramatically slowed down as they aggregate, making $v_{\phi}$, and therefore the active pressure, to drop (see eq. (\ref{eq:pa})).   %Thus, borrowing concepts from equilibrium thermodynamics, we call spinodal line the set of points in the Pe-$\phi$ plane defined by the onset of MIPS (see Fig. \ref{fig:phd} (a)).  
The  low-density branch of the binodal, $\phi_{\text{low}}$, is well below the density $\phi_{\text{n}}$ at which the dense phase appears. 
%{\bf IGN quizas estas dos fracciones volumicas se podrian indicar en el eje x de la figura 2c y 2d?. Asi quedaria mas visual y facil de identificar a que nos referimos con esta diferencia en lo que respecta a la forma de la curva de la presion} . 

Above $\text{Pe}_c$, and at  low densities, below the onset of MIPS, the equations of state collapse into a single curve  which is very well fitted by a first order virial expansion, $P=P_0(1+B_1\phi)$~\cite{Winkler2015}. For a system with confining walls,  $B_1$ decreases with Pe (see \cite{SM}) in consistence with the viewpoint that self-propulsion induces an effective particle attraction \cite{TailleurCates2008, Ginot2015, Marconi2015}. Surprisingly, for PBC, $B_1$ is independent of Pe, indicating that the activity-induced attraction saturates above Pe$_c$. 
 As we show below, this is due to the fact that in the absence of walls the system evolves along  a metastable branch as the density is increased and remains in its pure low-density phase above $\phi_{\text{low}}$, until it reaches the MIPS threshold, $\phi_{\text{n}}$, when a dense cluster  nucleates. %As we shall see in the following, the fact that the system is metastable between  $\phi_{\text{low}}$ and $\phi_{\text{n}}$ for Pe$>$Pe$_c$  explains the independence of $B_1$ on activity and the non-monotonic behaviour of the pressure.  

%Note that this is quite surprinsing since it means that the second virial coefficient remains unchanged as we move inside the coexistence region. 
Interestingly, the equations of state in confined and PBC geometries converge at large densities  to a value close to the expected coexistence pressure. 
%{\bf IGN: he cambiado el sentido de lo que querias decir?}
% to a value which is  rather close to the one at $\phi_{\text{low}}$, i.e. the expected coexistence pressure. %Our interpretation 
In  equilibrium, finite-size systems display  a pressure loop, which, in contrast to a van der Waals loop, is thermodynamically stable and due to the formation of an interface between the two phases
% once a drop as nucleated 
\cite{Mayer1965, Schrader2009}. Accordingly, $P(\phi)$ shows a peak  when the dense phase develops. In simulations, this discontinuous change is typically smoothed out by interface fluctuations.  The sharp decrease in $P$ for ABPs implies that interfacial fluctuations are considerably suppressed at high Pe, suggesting a large interface tension  (%in agreement with the phase diagram Fig. \ref{fig:phd}: 
AB-HDs might generate an interface between a dilute gas and a closed packed crystal). Hence, we can interpret the pressure drop for PBC systems as a finite size effect that should vanish in the thermodynamic limit. For finite systems the equation of state is affected both because of an interfacial  contribution to the pressure and  a shift in the location of nucleation~\footnote{For a first-order phase transition in equilibrium, the interface free energy density vanishes as $\Delta f\sim 1/\sqrt{N}$ and the nucleation point is shifted by $\phi_{\text{n}}\sim N^{-1/3}$ \cite{Mayer1965}.}.
These two effects are iconsistent with the data shown in Fig. \ref{fig:eos} (c). The active pressure in systems of size $N\gtrsim 4000$ converges above $\phi_{\text{n}}$, meaning that there is no interfacial finite-size contribution to the pressure after phase separation.  
It is less clear though whether the pressure jump is reduced when increasing $N$~\cite{Patch2016}. 
We push thermodynamic ideas even further, and claim that the system sizes used so far are not large enough to observe how the nucleation  approaches the binodal while the pressure drop vanishes. The idea behind this claim is that MIPS involves a large critical nucleus: our systems turn out to be too small to phase separate close to the binodal %between $\phi_{\text{low}}$ and $\phi_{\text{n}}$ 
and thus remain \emph{metastable} in a sub-region of the coexistence region~\footnote{Small systems should rather be thought in terms of clusters as discussed in \cite{McGinty1973, Lee1973, Reguera2003} for equilibrium systems}.
The convergence of $\phi_{n}$ towards the binodal as the system is made larger is due to the fact that the probability to spontaneously form a cluster of size $m>m_c$, (where $m_c$ is the critical nucleus size)  increases with system size. 
However, if  $N\gg m_c$ is not guaranteed, then we cannot  expect to observe how $P(\phi)$ approaches its infinite-size behavior.  
%If the system sizes used are not large enough to be able to form a nucleus, i.e. we can't guarantee $N\gg m_c$, then we do not expected to observe how $P(\phi)$ approaches its thermodynamic behaviour.  
In the following we show this important claim.% by two different routes.
%{\bf IGN: estoy confundido, queremos enfatizar que el valor del maximo decrece (peak) o que la caida disminuye (diferencia en tre el maximo y minimo en $\phi_n$?}
%: we quench the system with PBC from an initially phase separated state and we compute the pressure in the presence of confining walls. 

%much larger system sizes than the ones used so far would be needed in order 
%As the volume of the system is reduced particles aggregate to from clusters, and one of them must grow above a critical nucleus size in order to have phase coexistence.  with an effective barrier associated to it and related to the interface tension must be overcomed, giving rise to a 
%
%vanish  in the thermodynamic limit.
%
 %should reduce with the system size. As shown in Fig. \ref{fig:eos} (d) and discussed in \cite{Patch2016}, the simulations do not clearly show a loop vanishing in the thermodynamic limit. This might be due to 
 %This finite-size effect is due to the free energy penalty associated to the spontaneous formation of an interface from an initial homogenous state and, as such, it should vanish in the thermodynamic limit. 

In order to probe metastability, we initialize the open system deeply in the coexistence region. As we expand the system 
%, approaching the binodal from above, 
 nucleation  is avoided
%it does not have to overcome any nucleation barrier 
since an interface between a dense and low density phase is already present from the very beginning of the simulation. We performe such a  \emph{low-$\phi$ quench} at fixed Pe by letting AB-HDs evolve from a steady-state at $\phi=0.50$, $\text{Pe}=30$ to lower packing fractions. The equation of state we obtain is depicted in blue in Fig. \ref{fig:eos} (d). We find that: (i) the system remains phase separated at much lower densities than the nucleation point found by high-Pe quenches (see   Fig. \ref{fig:nuc} (a)); (ii) the pressure remains roughly constant down to $\phi\approx\phi_{\text{low}}$. We thus found \emph{hysteresis} around MIPS, a typical signature of  first-order phase transitions. 

%Starting from a homogeneous state, the system is not able to phase separate due to a large nucleation barrier that can not be overcome using systems of this size and remains in the \emph{metastable region}. 
 %, such that the system does not have to deal with the interface formation 
%The stabilization of thermodanamically stabel phase separated states in equilibrium finite-size systems also ,  
%metastable region between the binodal and spinodal lines and it is not able to reach the expected    
%Our interpretation is that the system is in a metastable branch
% , and both pressures converge to the same value at 

Accordingly to classical nucleation theory (CNT),  in the absence of a preferential site (homogeneous nucleation), phase separation can only be triggered by a rare event: the spontaneous formation of a critical nucleus of size larger than $m_c\propto \gamma/\Delta G_{\text{homo}}$ (where $\gamma$  is the tension  interface and $\Delta G_{\text{homo}}$ their free energy difference). 

%In  classical nucleation theory (CNT), the free energy associated with the  formation of a dense drop in a saturated gas is expressed as a sum of bulk and surface terms \cite{Oxtoby1992}. In the absence of a preferential site (homogeneous nucleation) the CNT free energy reads $\Delta G_{\text{homo}}=-S(G_{\text{l}}-G_{\text{h}})+\gamma_{\text{lh}} l $, where the sub-index ${\text{l}}$ and ${\text{h}}$ denote the low- and high-density phases, respectively, $S$ is the surface of the dense phase, $\gamma_{\text{lh}}$ is the tension of the interface and $l$ its length.  The competition between  bulk and interface free energies results in an energy barrier $\Delta G_{\text{homo}}^*$.  Thus, phase separation can only be triggered by a rare event: the spontaneous formation of a critical nucleus of size larger than $m_c\propto \gamma_{\text{lh}}/(G_{\text{l}}-G_{\text{h}})$. 
%{\bf IGN: este parrafo es necesario? parece un resumen de CNT pero las magnitudes que introducimos no se usan, excepto por la expresion de $m_c$. Es esto lo que te interesa remarcar? si es asi, quizas se podria reducir algo este parrafo y focalizarlo en este aspecto}

Since our system is intrinsically out-of-equilibrium, CNT cannot be directly applied. However, borrowing ideas from equilibrium systems, Redner et al.  have developed a theory analogous to CNT to describe the kinetics of phase separation in ABP \cite{Redner2016} which provides a good theoretical description of several simulation results \cite{Richard2016}. 
%For example, it predicts $m_c\approx 5000$ at Pe$=30$ and $\phi=0.30$ for AB-WCAD with $\epsilon=1$ in our units. 
Within this framework, the critical nucleus $m_c\propto \phi_{\text{cp}}/\ln^2(\text{Pe}\, \phi_{\text{low}})$ where $\phi_{\text{low}}$ is predicted by the theory. To make a comparison with simulations, an expression $\phi(A,\phi_{\text{high}}, \phi_{\text{low}})$ for the average packing fraction  in terms of the theoretical predictions is established, where $A$ is a structural parameter completely determined by the cluster size distribution $\mathcal{P}_m$ (see eq. (11) in \cite{Redner2016}). 
%The parameter $A$ contains the information about the structure of the system in terms of clusters and it is completely determined by the cluster size distribution $\mathcal{P}_m$. 
The nucleation barrier is thus controlled by two main ingredients: the location of the binodals and the structure of the clusters in the metastable region. We computed $\mathcal{P}_m$ for AB-HD and AB-WCAD with different stiffness and found a roughly identical distribution  (see Fig. \ref{fig:nuc} (c)), meaning that the nucleation barrier is mainly controlled by the location of the binodals. 
For stiffer potentials the structural difference between the coexisting phases is more severe. In the hard-disk limit a nearly perfect crystal coexists with a very dilute gas in the high activity regime, pushing the binodals to its extreme values, $\phi_{\text{low}}\to0$ and $\phi_{\text{high}}\to\phi_{\text{cp}}$. Therefore, the critical nucleus is expected to be very large at high Pe. As shown in Fig. \ref{fig:nuc} (d), the crystalline order of the dense phase is suppressed by softening the particles.  In turn, $\phi_{\text{low}}$ is larger for softer potentials such that $m_c$ is made smaller than in the hard limit case, thus qualitatively explaining the reduction of the pressure drop for softer disks (see Fig. \ref{fig:eos} (b)).
To be specific, Redner et al. predict $m_c\approx 5000$ at Pe$=30$ and $\phi=0.30$ for AB-WCAD with $\epsilon=1$ in our units (see Fig. 2 in \cite{Redner2016}). This value lies quite close to the binodal in their case while for AB-HD it falls deep into the coexistence region since the low-density branch of the binodal is shifted to much lower packing fractions in the hard disk limit.  We did not intend to make a direct quantitative comparison between the theory and our simulations, but, at this level, we are able to insure that for hard ABPs, MIPS features a large nucleation barrier that discourages attempts to observe how the system escapes from metastability with BD simulations. To give further support to this idea we turn now our attention into the system in presence of hard walls.

%So, from one hand, the harder the particles are, the more pronounced is the structural difference between the coexisting phases: in the hard-disk limit a nearly perfect crystal coexists with a very dilute gas, giving rise to a large effective interfacial tension $\gamma_{\text{lh}}$. From another hand, this extreme difference between the coexisting phases in the hard disk regime ($\epsilon\gg 1$)

Confinement facilitates nucleation because of  \emph{wetting}. %of particles on the walls. % surfaces act as natural nucleation sites. 
As shown in Fig. \ref{fig:nuc} (b), self-propelled particles accumulate at walls which thus act as natural 
nucleation seeds~\cite{Fily2014, Marconi2015, Speck2016}.
The free energy associated with (heterogeneous) nucleation in a confined system is 
$\Delta G_{\text{het}}=F(\alpha)\, \Delta G_{\text{homo}}$, where  $\ F(\alpha)=(2+\cos\alpha)(1-\cos\alpha)^2/4$ is a geometric function of the contact angle $\alpha$ between the wall and the dense phase.  The adsorption of ABPs into layers give $\alpha=0$ (pure wetting) and therefore, by extension of CNT, a vanishing nucleation barrier $\Delta G_{\text{het}}=0$. This equilibrium-like description is consistent with the  absence of a pressure drop in the presence of walls and confirming our overall interpretation of the equations of state across MIPS    (see Fig. \ref{fig:eos} (b)). 

We have carefully examined the pressure of ABPs using different potentials, topologies and preparation protocols in order to show that the equations of state are fully consistent with the classical (equlibirum) first-order phase transition scenario. %, reconciling numerical and theoretical results. 
The equations of state of ABP do not exhibit any fundamental difference to an equilibrium system showing phase coexistence, besides: (i) the absence of a Maxwell construction on $P$; (ii) the extreme structural difference between the two coexisting phases, giving rise to a large nucleation barrier. %In this respect, the use of infinitely hard disks has proven most helpful. 
Overall, our work results on a systematic way to interpret the equations of state of ABP %obtained in different situations 
using equilibrium-like concepts. We  quantitatively confirm the debated analogy between MIPS and equilibrium phase separation,  and, as such our work should represent an important step towards the construction of an effective thermodynamic description of active systems.%  and be useful to exploit and control aggregation phenomena in microswimmer suspensions. 

\paragraph*{Acknowledgments} 
DL warmly thanks Ran Ni, Gabriel Redner, Alexandre Solon, Thomas Speck, Caleb Wagner and Thomas Voigtmann for useful exchanges, and in particular Julien Tailleur for his comments and suggestions. DL acknowledges funding from a Marie Curie IE-Fellowship (G.A. no 657517).   I.P. acknowledges MINECO and DURSI for financial support under projects FIS2015-67837- P and 2014SGR-922, respectively.

\bibliographystyle{apsrev4-1}

\bibliography{EOS}

\end{document}